# Adiabatic population transfer of dressed spin states with quantum optimal control

Shu-Hao Wu, Mayra Amezcua, and Hailin Wang

Department of Physics, University of Oregon, Eugene, Oregon 97403, USA

Abstract

We report theoretical studies of adiabatic population transfer using dressed spin states. Quantum optimal control using the algorithm of Chopped Random Basis (CRAB) has been implemented in a negatively charged diamond nitrogen vacancy center that is coupled to a strong and resonant microwave field. We show that the dressed spin states are highly effective in suppressing effects of spin dephasing on adiabatic population transfer. The numerical simulation also demonstrates that CRAB-based quantum optimal control can enable an efficient and robust adiabatic population transfer.

## I. INTRODUCTION

Stimulated Raman adiabatic passage (STIRAP) has been extensively used for quantum control of three-level systems, in particular, Λ-type systems that feature two lower spin states with robust spin coherence[1-3]. In STIRAP, the quantum system evolves adiabatically, following a dark instantaneous eigenstate, which is a coherent superposition of the two lower states controlled by two external driving pulses with a counterintuitive pulse sequence. The fidelity of the STIRAP process is limited by the finite lifetime of the spin coherence and, for many solid-state spin systems, spin dephasing induced by a slowly fluctuating environment. Modifications to STIRAP have been pursued in order to speed up the relatively slow adiabatic passage[4-6]. One approach uses an auxiliary pulse to induce coupling between the two lower states[7,8]. This counter-diabatic coupling can be designed to cancel precisely nonadiabatic transitions between the instantaneous eigenstates induced by the two driving pulses, leading to a net transitionless driving. Another approach uses specially-designed temporal shapes for the two driving pulses to achieve superadiabatic transitionless driving (SATD). This approach can be implemented in the limit of either zero dipole detuning or large dipole detuning[9-11]. In addition, systematically-corrected control pulses using Magnus expansion techniques have also been proposed for the achievement of transitionless driving[12].

Both conventional STIRAP and SATD STIRAP have been applied to negatively-charged nitrogen vacancy (NV) centers in diamond[13,14], a solid state spin system that has recently emerged as a promising qubit system for quantum information processing[15-18]. In spite of these successful experimental demonstrations, the transfer efficiency observed in both experiments are far from ideal. This is by no means surprising. Although spin decoherence time exceeding 1 ms has been observed in isotopically purified diamond samples[19], NV centers in diamond samples with natural abundance of $^{13}C$ are subject to slow fluctuations of the $^{13}C$ nuclear spins. This nuclear spin bath leads to $T_2^*$, the effective spin dephasing time, on the order of 1 μs. This spin dephasing process can be suppressed with dynamical decoupling techniques[20]. The sequence of π-pulses used in dynamical decoupling, however, is incompatible with adiabatic passage. Alternatively, the spin dephasing process can be suppressed with the use of dressed spin states, using microwave fields to dress the electron spin states[21-23].

In this paper, we report theoretical studies of adiabatic population transfer using dressed spin states. To avoid the nonadiabatic excitation of the upper state in the Λ-type three-level system, we have implemented the adiabatic population transfer with a large optical detuning. The implementation is based on the quantum optimization technique of Chopped Random Basis (CRAB)[24,25], a recently developed numerical algorithm for quantum optimal control[26]. We show that for diamond with natural $^{13}$C abundancy, transfer efficiency exceeding 97% can be achieved with the CRAB-optimized temporal pulse lineshapes and with the use of the dressed spin states. Overall, these studies demonstrate that dressed spin states are highly effective in overcoming detrimental effects of spin dephasing in the adiabatic passage process and that the CRAB algorithm can provide a simple and yet powerful technique for quantum optimal control of the adiabatic passage.

## II. THEORETICAL RESULTS

**a) Λ-type systems and dressed spin states**

For a NV center in diamond, either the $E_y$ or $E_x$ excited state can serve as the upper state, |e>, and the $m_s$=0 and the $m_s$=-1 (or $m_s$=1) states can serve as the two lower spin states of a Λ-type three-level system, as shown schematically in Fig. 1a. The dipole optical transition between $E_y$ (or $E_x$) state and the $m_s$=-1 state can be enabled by strain induced excited-state mixing or the excited state level anti-crossing, as demonstrated in earlier studies[27,28]. Coherent population trapping (CPT) in this Λ-type system has already been demonstrated in a number of earlier experiments[27-29]. To use dressed spin states to suppress effects of spin dephasing, we apply a strong and resonant microwave (MW) field, with frequency $\omega_B$, between the two lower spin states. The corresponding Hamiltonian in the rotating frame is described by

$$\bar{H}_{mw} = \hbar\delta|-1\rangle\langle-1| + \hbar\frac{\Omega_0}{2}(|0\rangle\langle-1| + |-1\rangle\langle0|) \tag{1}$$

where $\delta$ is the bath-induced energy fluctuation of the |−1> state and $\Omega_0$ is the Rabi frequency for the MW driving field. The eigenstates of this Hamiltonian are the semiclassical dressed states given by

$$|+\rangle = sin\theta_0|0\rangle + cos\theta_0|-1\rangle \tag{2a}$$

$$|-\rangle = cos\theta_0|0\rangle - sin\theta_0|-1\rangle \tag{2b}$$

where $\theta_0 = tan^{-1}[\Omega_0/(\sqrt{\Omega_0^2+\delta^2}+\delta)]$. The corresponding eigenvalue is $E_\pm = \pm\hbar\sqrt{\Omega_0^2+\delta^2}/2$. The energy of the dressed spin states can be immune to the fluctuations of the nuclear spin bath if $\Omega_0 \gg |\delta|$. For the numerical calculations, we have set $\Omega_0/2\pi=50$ MHz. These dressed spin states will be used as the two lower states of the new Λ-type system, as illustrated schematically in Fig. 1b.

For adiabatic passage between the two dressed spin states, we start with an initial state,

$$|\psi>_i = (|0>-|-1>)/\sqrt{2}. \tag{3a}$$

To prepare the system in this initial state, we first initialize the NV center in the $m_s$=0 state and then use a π/2 pulse to drive the NV into $|\psi>_i$, as shown schematically in Fig. 1c. The final state or the target state is given by

$$|\psi(t)>_f = (|0>+e^{-i\omega_B t}|-1>)/\sqrt{2}. \tag{3b}$$

Following the adiabatic passage process along with a suitable waiting period, $|\psi(t)>_f$ can be driven to |−1> with a π/2 pulse for the experimental state detection, as indicated in Fig. 1c.

**b) Adiabatic population transfer**

For adiabatic population transfer, the overall Hamiltonian is given by $H = H_0 + H_{mw} + V$, where $H_0 = \hbar\omega_e|e><e|+\hbar(\delta+\omega_B)|-1><-1|$ is the Hamiltonian for the three-level system and $H_{mw}$ is the microwave interaction Hamiltonian in the Schrodinger picture. The optical interaction Hamiltonian is given by

$$V = \frac{\hbar\Omega_s(t)}{2}(|-1\rangle\langle e|e^{i\omega_s t}+h.c.) + \frac{\hbar\Omega_p(t)}{2}(|0\rangle\langle e|e^{i\omega_p t}+h.c.), \tag{4}$$

where $\Omega_s(t)$, $\Omega_p(t)$, $\omega_s$, and $\omega_p$ are the Rabi frequencies and optical frequencies for the two optical driving fields, respectively. While a set of density matrix equations with the dressed-state basis can be used, these equations, which include unusual decay terms[30], are more difficult to use than the much simpler density matrix equations for the bare states. In this regard, we have carried out numerical calculations of the adiabatic passage process using the Lindblad master equation for the bare states, which includes explicitly interactions of both MW and optical fields as well as the population decay rate for |e>, Γ, and the spin decoherence rate, $\gamma_s$. The Lindblad master equation used is given by

$$\dot{\rho}(t) = -\frac{i}{\hbar}[H, \rho(t)] + \sum_k [L_k \rho(t) L_k^\dagger - \frac{1}{2}\{L_k^\dagger L_k, \rho(t)\}], \tag{5}$$

where $L_0 = \sqrt{\Gamma/2}\,|0\rangle\langle e|$, $L_1 = \sqrt{\Gamma/2}\,|-1\rangle\langle e|$ and $L_3 = \sqrt{\gamma_s}\,(|-1\rangle\langle -1| - |0\rangle\langle 0|)$ are the jump operators. Note that the Raman resonant or two-photon resonant condition (with $\delta$=0) for the adiabatic passage of the relevant dresses spin states is given by $\omega_p - \omega_s = \omega_B + \Omega_0$, where the formation of the MW-driven dressed states leads to a corresponding shift in the Raman resonant condition.

An important constraint in the adiabatic population transfer between dressed spin states is that $\Omega_{eff}$, the effective Rabi frequency between the two spin states induced by the external optical fields, needs to be small compared with $\Omega_0$, the MW Rabi frequency used for creating the dressed states. This is because power broadening of the effective two-level transition can lead to undesirable excitations of nearby dressed states. In addition, in the limit of large dipole optical detuning, the effective optically-induced decay rate due to the nonadiabatic excitation of the excited state is $(\Omega/2\Delta)^2 \Gamma$, where $\Omega$ is the relevant optical Rabi frequency and $\Delta$ is the dipole detuning, with $\Gamma/2\pi$ =14 MHz for NV centers in diamond[31]. Given the above two constraints, we set the maximum optical Rabi frequency, $\Omega_s/2\pi$ and $\Omega_p/2\pi$, to be $\sqrt{2}$ *100 MHz, along with $\Delta/2\pi$>1.5 GHz.

**c) CRAB and far off-resonant SATD**

For adiabatic population transfer with CRAB-based quantum optimal control, we optimize the transfer efficiency from the initial to the target states by varying the temporal pulse lineshapes of the two external optical fields, with a fixed overall duration for the transfer process and a fixed dipole optical detuning. We also constrain the optimization process on a maximum optical Rabi frequency. The numerical implementation of the CRAB algorithm has been carried out with the use of Quantum Toolbox in Python (QuTiP), an open-source software for simulating the dynamics of open quantum systems.

For comparison, we have also carried out theoretical calculations using the far off-resonant SATD scheme. The far off-resonant SATD approximates the Λ-type three-level system as an effective two-level system in the limit of large dipole detuning. For a Λ-type three-level system with $\Delta = \omega_e - \omega_p = \omega_e - \omega_B - \omega_s$ and with $\Omega_0=\delta=0$, the effective Rabi frequency, $\Omega_{eff}$, between the two lower states is given by

$$\Omega_{eff} = \frac{\Omega_s(t)\Omega_p(t)}{2\Delta}. \tag{6a}$$

The effective detuning, $\Delta_{eff}$, for the two-level transition is

$$\Delta_{eff} = \frac{\Omega_p^2(t)-\Omega_s^2(t)}{4\Delta}. \tag{6b}$$

Solving the above two equations for $\Omega_s$ and $\Omega_p$, we can express the Rabi frequencies of the external optical fields in terms of the effective Rabi frequency and the effective detuning,

$$\Omega_p(t) = [2\Delta(\sqrt{\Omega_{eff}^2 + \Delta_{eff}^2} + \Delta_{eff})]^{1/2}, \tag{7a}$$

$$\Omega_s(t) = [2\Delta(\sqrt{\Omega_{eff}^2 + \Delta_{eff}^2} - \Delta_{eff})]^{1/2}. \tag{7b}$$

A counter diabatic coupling can be applied between the two lower states such that the overall system can remain in the instantaneous dark state, with the Rabi frequency for the counter diabatic coupling given by[10]

$$\Omega_a = \frac{2[\dot{\Omega}_p(t)\Omega_s(t)-\Omega_p(t)\dot{\Omega}_s(t)]}{\Omega_p^2(t)+\Omega_s^2(t)}. \tag{8}$$

For the far off-resonant SATD, the counter diabatic coupling can be implemented by a suitable modification of both $\Omega_{eff}$ and $\Delta_{eff}$. As shown in the earlier study[10], the modification is

$$\tilde{\Omega}_{eff}(t) = \sqrt{\Omega_{eff}^2(t) + \Omega_a^2(t)}, \tag{9a}$$

$$\tilde{\Delta}_{eff}(t) = \Delta_{eff}(t) + \frac{d}{dt}tan^{-1}(\frac{\Omega_a(t)}{\Omega_{eff}(t)}). \tag{9b}$$

Using Eq. 7, we can then derive the modified Rabi frequencies for the two external optical fields, $\tilde{\Omega}_s(t)$ and $\tilde{\Omega}_p(t)$.

Figure 2 shows the numerical calculations obtained for bare spin states ($\Omega_0=0$), for which we have assumed $\delta=0$ (no nuclear spin bath), a spin decoherence rate $\gamma_s/2\pi==1$ kHz, and $\Gamma/2\pi=14$ MHz. Figure 2a plots the temporal pulse lineshapes derived for the far off-resonant SATD. Figure 2b shows the temporal pulse lineshapes obtained with quantum optimal control using the CRAB algorithm. A striking feature of the CRAB temporal pulse lineshapes is that the overall pulse area for each optical pulse is near zero. Figure 2c shows the adiabatic transfer efficiencies obtained with the far off-resonant SATD and those obtained with the CRAB algorithm.

As shown in Fig. 2c, the transfer efficiency increases slightly with increasing detuning, which is expected since a large dipole detuning reduces nonadiabatic excitation of the upper state population. Furthermore, the transfer efficiency obtained with the CRAB algorithm is considerably greater than that obtained with far off-resonant SATD, since the far off-resonant SATD is not optimal for the detuning used for Fig. 2.

### d) Suppressing effects of spin dephasing with dressed spin states

Since the nuclear spin bath fluctuates at a relatively slow timescale, we assume that for the duration of the adiabatic population transfer, $E_{-1}$, which is the energy level of state $|-1\rangle$, remains unchanged. To describe spin dephasing induced by the nuclear spin bath, we let $E_{-1}$ take a value according to a normal distribution, with

$$E_{-1} = \hbar(\omega_B + \delta) = \hbar[\omega_B + N(0,\sigma^2)], \tag{10}$$

where $N(0,\sigma^2)$ is a normal distribution with mean=0 and variance=$\sigma^2$. For each $\sigma$, we ran a large number of simulations with the relevant detuning adjusted according to Eq. 10. Effects of spin dephasing are investigated with $\sigma/2\pi$ varied from 0 to 2 MHz. Numerical calculations were carried out for the dressed spin states with $\Omega_0/2\pi$=50 MHz and also for bare states (with $\Omega_0$=0), for which we calculate directly the transfer efficiency from state $|0\rangle$ to state $|-1\rangle$.

Figure 3a compares the adiabatic transfer efficiencies obtained for bare and dressed spin states as a function of $\sigma$ and with CRAB-based quantum optimal control. Identical optical temporal lineshapes are used for both the dressed state and bare spin state calculations. Similar to Fig. 2b, the duration of the population transfer is set to 0.72 μs. A maximum Rabi frequency of $\Omega/2\pi = \sqrt{2}*100$ MHz is used for the dressed spin states, while a maximum Rabi frequency of $\Omega/2\pi = 100$ MHz is used for the bare spin states. This is because in order to have the same $\Omega_{eff}$ for the bare and the dressed spin states, $\Omega_s$ and $\Omega_p$ used for the dressed spin states needs to be a factor of $\sqrt{2}$ of those used for the bare spin states.

For the bare spin states, the transfer efficiency displays large fluctuations as $\sigma$ increases, which is expected because of the corresponding fluctuations in the relevant Raman detuning. In comparison, the transfer efficiency remains nearly 98% for the dressed spin states, demonstrating nearly complete suppression of the effects of spin dephasing on the adiabatic population transfer. Figure 3b shows the histogram of a theoretical simulation, for which we took $\sigma/2\pi=2$ MHz and carried out 1000 calculation runs with $E_{-1}$ taking a value according to the normal distribution given in Eq. 10. As shown in the histogram, the population transfer efficiency for the bare states exhibits a broad distribution ranging from 0.8 to greater than 0.99. In contrast, for the dressed spin states, the distribution of the population transfer efficiency is characterized by a sharp peak centered near 0.98. The width of the peak is less than 1%, which again confirms the nearly complete suppression of the effects of spin dephasing on adiabatic population transfer by the dressed spin states.

It should be noted that the dressed spin states exhibit a slightly greater nonadiabatic excitation of the excited states than the corresponding bare states, as shown in Fig. 3c. This is due to the fact that greater optical Rabi frequencies are used for the dressed spin states. In addition, the excited state population is nearly independent $\sigma$ for the dressed states, but shows some variation with increasing $\sigma$ for the bare states, as shown in Fig. 3c.

The theoretical results shown in Figs. 3a and 3b assume perfect state initialization and readout for the dressed spin states. Both of these steps, however, are also subject to spin dephasing. Figure 4 shows the histogram of the transfer efficiency obtained under the same condition as that used for Fig. 3b, except that effects of spin dephasing on the state initialization and readout are included in Fig. 4a and Fig. 4b, respectively. For Fig. 4c, effects of spin dephasing on the entire process including state initialization, transfer, and readout are included. The average efficiency obtained from Figs. 4a, 4b, and 4c are 97.4%, 97.43%, 97.06%, respectively. In comparison, the average efficiency obtained from Fig. 3b is 97.72%.

## III. CONCLUSION

In conclusion, theoretical studies using NV centers in diamond as a model system demonstrate that dressed spin states can be a highly effective approach for suppressing or circumventing effects of spin dephasing in adiabatic population transfer. The numerical simulations also show that the simple CRAB algorithm is a powerful tool for optimal quantum control, enabling efficient and robust adiabatic population transfer. These results should stimulate further experimental efforts on adiabatic population transfer in solid state spin systems.

**ACKNOWLEDGEMENTS**

This work is supported by a grant from ARO MURI and by NSF under grants No. 1719396.

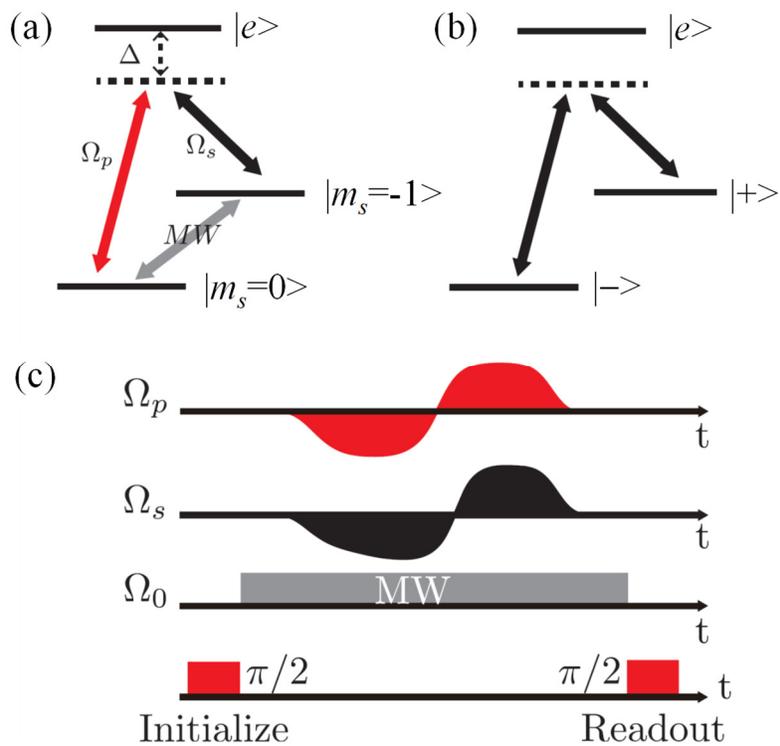

Fig.1 (color online). (a) Schematic of a Λ-type three level system in a NV center, for which the two lower states can also couple to a strong and resonant microwave field. (b) Schematic of the resulting Λ-type three level system with the dressed spin states as the two lower states. (c) Schematic of the pulse sequence for adiabatic population transfer of dressed spin states. The π/2 microwave pulses are used for the preparation and readout of dressed spin states.

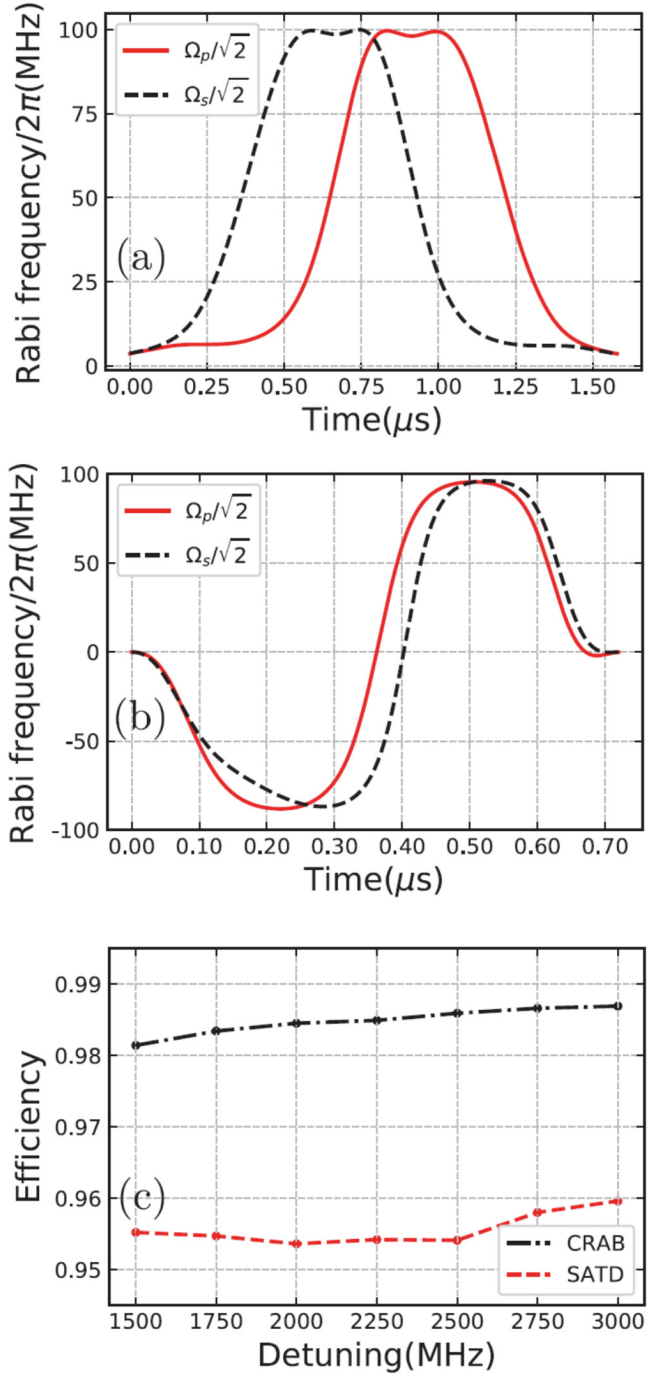

Fig. 2 (color online). (a) Temporal pulse lineshapes of the two external optical fields for the far off-resonant SATD, with dipole detuning Δ=3 GHz. (b) Temporal pulse lineshapes of the two external optical fields optimized with the CRAB algorithm, with Δ=3 GHz. (c) Adiabatic population transfer efficiency of dressed spin states as a function of $\Delta/2\pi$, obtained with the far off-resonant SATD and CRAB schemes. No effects of spin dephasing due to the nuclear spin bath are included.

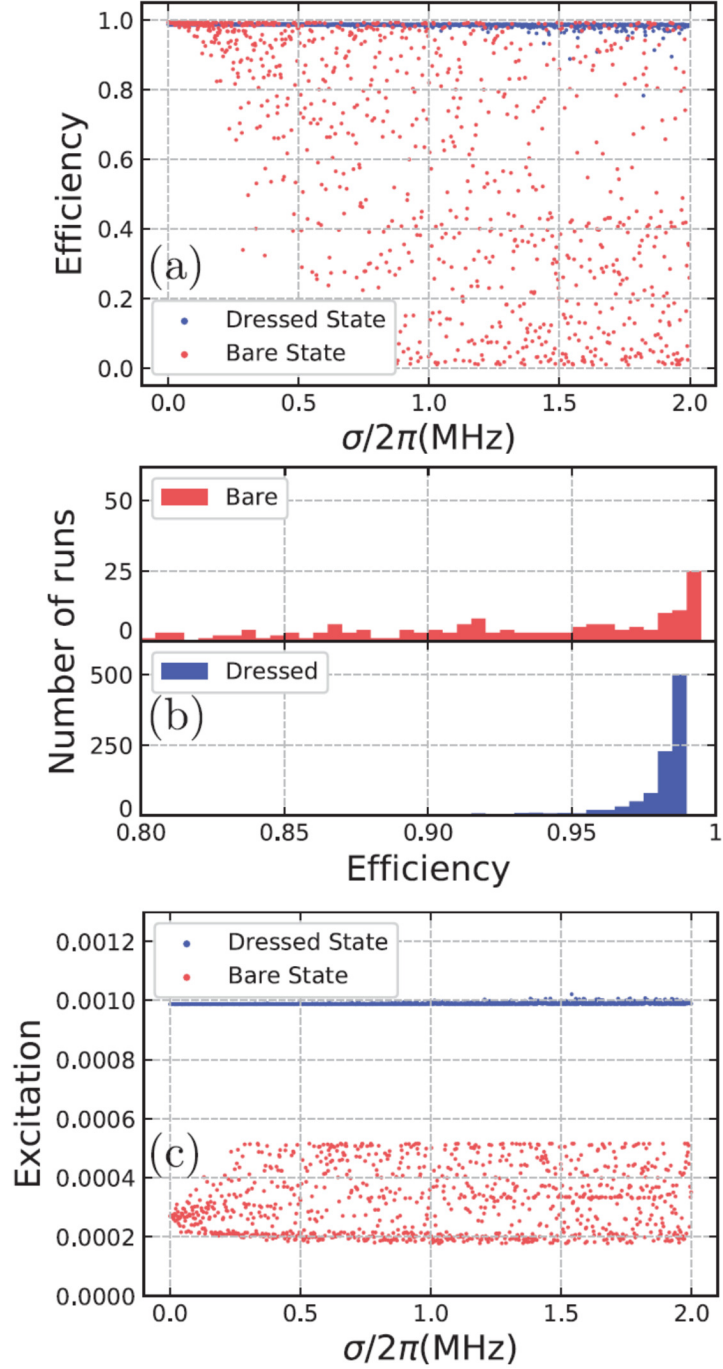

Fig. 3 (color online). (a) Adiabatic population transfer efficiency for dressed spin states and bare spin states obtained with the CRAB algorithm and with Δ=3 GHz, as σ/2π is varied from 0 to 2 MHz. For each data point, $E_{-1}$ takes a value according to the normal distribution given in Eq. 10. (b) The distribution of the adiabatic population transfer efficiency for dressed spin states and bare spin states (with Δ=3 GHz). A total number of 1000 simulation runs with σ/2π=2 MHz are plotted for each distribution. (c) Nonadiabatic excitation of the excited state for the simulation runs shown in (a).

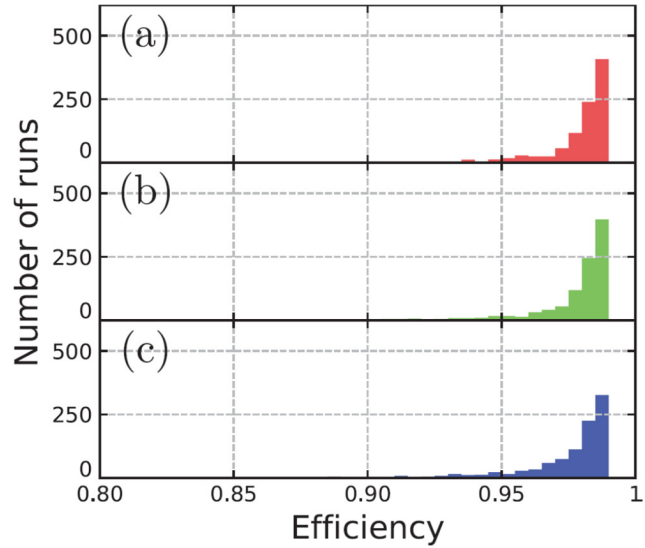

Fig. 4 (color online). The distribution of the adiabatic population transfer efficiency for dressed spin states. A total number of 1000 simulation runs with $\sigma/2\pi=2$ MHz are plotted for each distribution. From the top to bottom, effects of spin dephasing are including in initialization and transfer, in readout and transfer, and in initialization, transfer, and readout, respectively. Other conditions are the same as those used in Fig. 3b.